\documentclass[reprint, prd, aps, preprintnumbers]{revtex4-1}
\usepackage{amsmath}

\usepackage{amssymb}
\usepackage{graphicx}
\usepackage{color}
\usepackage{soul} %

\providecommand{\update}[1]{#1}

\renewcommand{\vec}[1]{\boldsymbol{#1}}

\providecommand{\set}[1]{\mathcal{#1}}

\providecommand{\HERON}{Heron} %
\providecommand{\SEOBNR}{SEOBNR}
\providecommand{\PHENOM}{IMRPhenomP}
\providecommand{\dd}[1]{\text{d}#1}

\providecommand{\transpose}[1]{{#1}^{\mathsf{T}}}

\usepackage{acronym}

\providecommand{\GP}{\ac{GP}}

\providecommand{\trainingpoints}{\mathcal{X}}
\providecommand{\trainingobservations}{\mathcal{Y}}
\providecommand{\trainingdata}{(\trainingpoints, \trainingobservations)}
\providecommand{\kernel}[1]{\mathsf{#1}}
\providecommand{\SE}{\kernel{SE}}

\providecommand{\numbertrainingwaveforms}{$132$}

\graphicspath{{figures/}}

\begin{document}

\newacro{NR}{Numerical Relativity}%
\newacro{GW}{gravitational wave}%
\newacro{BBH}{Binary black hole}%
\newacro{LSC}{LIGO Scientific Collaboration}%
\newacro{GPR}{Gaussian process regression}%
\newacro{PE}{parameter estimation}%
\newacro{GP}{Gaussian process}%
\newacro{LOO}{leave-one-out}
\newacro{HODLR}{Hierarchical off-diagonal low rank}

\title{A Precessing Numerical Relativity Waveform Surrogate Model for Binary Black Holes: A Gaussian Process Regression Approach}
\author{D. Williams}
\email{daniel.williams@glasgow.ac.uk}
\affiliation{SUPA, University of Glasgow, G12 8QQ}
\author{J. Gair}
\affiliation{University of Edinburgh}
\author{I. S. Heng}
\affiliation{SUPA, University of Glasgow, G12 8QQ}
\author{J. A. Clark}
\affiliation{Georgia Institute of Technology}
\author{B. Khamesra}
\affiliation{Georgia Institute of Technology}
\date{\today}
\preprint{DCC:LIGO-P1800128}

\begin{abstract}
  Gravitational wave astrophysics relies heavily on the use of matched filtering both to detect signals in noisy data from detectors, and to perform parameter estimation and tests of general relativity on those signals.
  Matched filtering relies upon prior knowledge of the signals expected to be produced by a range of astrophysical systems, such as binary black holes.
  These waveform signals can be computed using numerical relativity techniques, where the Einstein field equations are solved numerically, and the signal is extracted from the simulation.
  Numerical relativity simulations are, however, computationally expensive, leading to the need for a surrogate model which can predict waveform signals in regions of the physical parameter space which have not been probed directly by simulation.
  We present a method for producing such a surrogate using Gaussian process regression which is trained directly on waveforms generated by numerical relativity.
  This model returns not just a single interpolated value for the waveform at a new point, but a full posterior probability distribution on the predicted value.
  This model is therefore an ideal component in a Bayesian analysis framework, through which the uncertainty in the interpolation can be taken into account when performing parameter estimation of signals.

\end{abstract}

\maketitle

\section{Introduction}
\label{sec:introduction}

The first detection of gravitational waves in September 2015 was the result not only of advanced detector development, but also the development of data analysis techniques which were capable of detecting and characterising weak signals in noisy data.
The most sensitive of these techniques rely on \emph{matched filtering} to identify signals, and these techniques are most effective when accurate and efficient waveform models are available to produce template banks.

The production of high-accuracy waveforms is possible thanks to advances in the field of \ac{NR}, in which the full set of Einstein equations are solved numerically.
This can be done reliably for the low-mass compact binary systems of interest to the current generation of ground-based gravitational wave observatories,
however these simulations are computationally expensive, and can require  thousands of CPU hours to run in situations where the mass ratios and spins of the black holes are small.
A simulation of a full 350-cycle gravitational waveform spanning the entire advanced LIGO band has been produced~\cite{2015arXiv150204953S}, however this required several months of high-performance computing to complete~\cite{2016CQGra..33l5025D}, despite employing numerous techniques to reduce wall-clock computation time.
As a result only around 1000 \ac{NR} waveforms are available, and most of these are much shorter than 350 cycles long.
\ac{BBH} coalescences are described by a number of physical parameters:
the ratio of the two component black holes' masses, $q$;
the vector of each component's spin, $\vec{s}_1$ and $\vec{s}_2$;
and the time, $t$, relative to a fixed reference time, for example the time of coalescence of the binary.

This results in a parameter space with eight dimensions which is very sparsely sampled.
As a result, \ac{NR} waveforms alone are not a practical way to form the tempate banks required for precise signal parameter estimation.
In addition, the high cost of producing new simulations is unlikely to significantly change this situation in the near future.

To overcome this problem there have been significant efforts to inform analytical models of non-spinning black hole coalescences with the results of \ac{NR} simulations of spinning systems to produce an analytical, phenomenological approximant which can be rapidly evaluated.
There are two major implementations of analytical models which are calibrated against \ac{NR}-derived waveforms, the Phenom and \SEOBNR{} families of approximants.
The Phenom family have developed from \texttt{IMRPhenomA}~\cite{2008PhRvD..77j4017A}, which was capable of producing waveforms for non-spinning binaries, through to \texttt{IMRPhenomD}~\cite{2016PhRvD..93d4007K}, which models spinning, non-precessing binaries.

The Phenom family of waveforms has been developed to incorporate support for precessing systems through the \PHENOM{} codes; the latest edition of this is \texttt{IMRPhenomPv3}~\cite{2018arXiv180910113K}, although in this work we will make use of the slightly older \texttt{IMRPhenomPv2}~\cite{2016PhRvD..93d4007K}, which has extensive support within the \texttt{pyCBC}~\cite{2016CQGra..33u5004U,alex_nitz_2017_883086,2014PhRvD..90h2004D} library used in the preparation of this work.
This is composed of a post-Newtonian approximation to the inspiral period of the waveform, and a phenomonological ansatz for the merger and ringdown periods.
The approximant is calibrated against 19 \ac{NR}-derived waveforms to produce a model which has a low mismatch (defined in equation \ref{eq:mismatch}) with the calibration data.

The \SEOBNR{}  family provide an alternative approach to that taken by the \PHENOM{} models, using an effective one-body approach~\cite{1999PhRvD..59h4006B,2000PhRvD..62f4015B,2009arXiv0906.1769D} to map the dynamics of a binary into those of a single test particle in a deformed Kerr metric.
In contrast to the piece-wise approach to building the waveform from the inspiral, merger and ringdown of the \PHENOM{} models, the \SEOBNR{}  models construct the waveform through a single process~\cite{2011PhRvD..84l4052P}.
A number of models based on the effective one-body approach exist, ranging from \texttt{EOBNR} which model non-spinning systems~\cite{2007PhRvD..76j4049B,2011PhRvD..84l4052P} to the \SEOBNR{} families of model, which can model spinning systems~\cite{2012PhRvD..86b4011T,2014PhRvD..89f1502T,2017PhRvD..95d4028B}, and precession effects~\cite{2014PhRvD..89h4006P}.
Similarly to \texttt{IMRPhenom}, these models are calibrated against \ac{NR} waveforms: for \texttt{SEOBNRv3} five waveforms are used for this calibration.

These models can be evaluated quickly, and are thus suitable for the rapid parameter estimation tasks required for the detection and characterisation of gravitational waves.
However, both the Phenom and \SEOBNR{}  models are affected by systematic uncertainties which are difficult to quantify in regions of the \ac{BBH} parameter space which are not calibrated against \ac{NR} simulations.

The NRSur family of surrogate models, developed by Blackman \emph{et al.}~\cite{2015PhRvL.115l1102B,2017PhRvD..95j4023B,2017PhRvD..96b4058B} employ spline interpolation to waveforms generated by the SpEC \ac{NR} code.
The two analysis-ready versions of this model, \texttt{NRSur4d2s} and \texttt{NRSur7d2s} are capable of producing waveforms for systems with a mass-ratio $<2$ and an effective spin-parameter $< 0.8$. 
In contrast to phenomenological models, the NRSur models are currently capable of producing only a small number of cycles of the waveform, being limited by the length of the \ac{NR} waveforms off which they are conditioned.
Recent efforts have been made, however, to produce similar surrogate models which are conditioned on hybridised waveforms \cite{2018arXiv181207865V}.
The number of waveforms required to produce the surrogate model is also considerably larger than those required for the phenomenological models, with \texttt{NRSur7d2s} being conditioned on 744 \ac{NR} waveforms.

Efforts to account for the systematic uncertainty between \ac{NR} waveforms and waveforms produced by phenomenological models have been proposed in which the uncertainties are modelled by \ac{GPR}~\cite{2014PhRvL.113y1101M,2016PhRvD..93f4001M}.
This allows the uncertainty in the interpolation to be calculated from the posterior predictive distribution of the \ac{GPR}.
This probability distribution, derived from \ac{GPR} can be used to marginalise the likelihood of the observed \ac{GW} data over waveform uncertainty.
This approach was shown to provide a significant reduction in biases in \ac{PE} compared to using phenomenological methods with no attempt to account for the uncertainty~\cite{2014PhRvL.113y1101M,2016PhRvD..93f4001M}.

These previous efforts suggested using \ac{GPR} to model the difference between \ac{NR} waveforms and phenomenological models.
We propose to extend this approach by producing a model of the entire gravitational waveform using \ac{GPR} as a surrogate model conditioned only on numerical relativity simulations, without any reference to a phenomenological model.
In comparison to the NRSur families of surrogate, \ac{GPR} is capable of not only producing an approximant for the waveform throughout the parameter space, but also an uncertainty on that estimate.
We note that our model is not the first to attempt to predict \ac{BBH} waveforms using \ac{GPR}, but it is the most complete.
A previous model~\cite{2017zoheyrsurrogate} used \ac{GPR}, but this was conditioned on waveforms generated from the \texttt{IMRPhenomPv2} phenomenological approximant, and not \ac{NR} data, and is not capable of producing generically spinning waveforms.

\ac{GPR} is a Bayesian regression technique which relies on a \ac{GP} prior distribution.
A \GP{} can be considered as a prior over a space of functions, each of which are considered a potential fitting function to some set of data.
The \ac{GP} model assumes that the \update{values of the function} evaluated at a certain finite set of points \update{are draws} from a multi-variate Gaussian distribution.
The \ac{GP} prior is itself defined by a number of assumptions about the behaviour of these functions (e.g. their smoothness).
When the \ac{GP} prior model is conditioned on data from existing simulations (potentially allowing for uncertainties in each of the simulations), the resulting posterior provides a distribution of functions which could represent the true model. The mean of this posterior distribution can be used analogously to the single fitting function which is produced by more conventional regression techniques, while the variance of the distribution provides a measure of the goodness-of-fit.

The structure of this publication is as follows.
In section \ref{sec:gps} we explain the process used for the production of a waveform surrogate model, and the choice of covariance function for our model in section \ref{sec:covariancefunction}.
Our new model, named \HERON{} is introduced in section \ref{sec:heron}, with the waveforms used to train the model described in section \ref{sec:trainingdata}, and a discussion of the complications introduced by using a large quantity of data is provided in section \ref{sec:complexity}.
An overview of the testing procedures which we used to verify the output of the model is included in section \ref{sec:verification}, with both the results of these tests, and a number of example waveform outputs are presented in section \ref{sec:examples}.

\section{Gaussian Process Regression}
\label{sec:gps}

A \ac{GP} represents a distribution of potential functions which can explain a set of training data $\trainingdata$, composed of observations, $\trainingobservations$, made at locations, $\trainingpoints$, within the parameter space of the problem, such that the function values
\[ y = f(\vec{x}) \]
\update{(}for \update{each} $\vec{x} \in \trainingpoints, y \in \trainingobservations$\update{)}, are modelled as being drawn from a multivariate normal distribution.
As such, the \GP{} is fully characterised by its mean function, $\mu(\vec{x})$, and a covariance function, $k(x,x')$, which describes the similarity between two function values at two points in the parameter space.
A \GP{} can be defined with any positive-definite covariance function, the form of which encodes prior assumptions about  the data, for example its smoothness and stationarity.
Popular choices of covariance function include the squared exponential covariance functions, and Mat\'{e}rn covariance functions~\cite{2016PhRvD..93f4001M,Rasmussen:2005:GPM:1162254}.

It is common to assume the training samples have mean zero.
This causes the mean of the \ac{GP} to be zero outside the training set, which, while unphysical, is a reasonable assumption given a lack of data; within the region described by the training set the mean of the function is defined by the training data.
Making this assumption allows the mean squared properties of the data to be determined entirely through the covariance function.

When defining the covariance function for the \GP{} it is often desirable to specify a number of free hyperparameters, $\vec{\theta}$, which allow the properties of the covariance function (and hence the \GP{}) to be adapted based on the training data. Bayesian model comparison can be used to select the \GP{} which optimally describes the data, or to obtain a posterior distribution on appropriate values of the hyperparameters.
The log-probability that a given set of function values were drawn from a \GP{} with zero mean and a covariance matrix $K_{ij} = k(x, x')$ is

\begin{equation}
  \label{eq:logevidencegp}
  \log(p(\vec{y}| X)) = - \frac{1}{2} {\transpose{\vec{y}}} K^{-1} \vec{y} - \frac{1}{2} \log |K| - \frac{n}{2} \log 2\pi .
\end{equation}
\update{With $n$ the total number of points in the training data.} This quantity is normally denoted the \emph{log-evidence} or the \emph{log-hyperlikelihood}.
The model which best describes the training data may then be found by maximising the log-hyperlikelihood with respect to the hyperparameters, $\theta$ of the covariance function.

Once the \GP{} has been conditioned on the training data
and the optimal covariance function identified through model
comparison, it is possible to exploit it as a predictive tool,
allowing the interpolation of function outputs between training
data. In order to make a prediction using the GP model we require a
new input point at which the prediction should be made, which is
denoted $x^*$. In order to form the predictive distribution we must
then calculate the covariance of the new input with the existing
training data, which we denote $K_{x, x^*}$, and the autocovariance of
the input, $K_{x^*, x^*}$. We then define a new covariance matrix,
$K^{+}$, which has the block structure
\begin{equation}
  \label{eq:blockK-plus-mat}
  K^+ =
  \begin{bmatrix}
    K_{x,x} & K_{x,x^*} \\ K_{x^*,x} & K_{x^*, x^*}
  \end{bmatrix},
\end{equation}
for $K_{x,x}$ the covariance matrix of the training inputs, and
$K_{x^*,x} = K_{x,x^*}^T$.

The predictive distribution can then be found as
\begin{equation}
  \label{eq:predictive-gp}
  p(\vec{y}^* | \vec{x}^*, \mathcal{D}) = \mathcal{N}(\vec{y}^* | K_{x^*,x} K_{x,x}^{-1} \vec{y}, K_{x^*, x^*} - K_{x^*,x}K^{-1}_{x,x} K_{x,x^*}),
\end{equation}
where $\mathcal{D}$ is the training data\update{, and $\mathcal{N}$ is the normal distribution}.

Equation \ref{eq:predictive-gp} emphasises the value of the GP
approach to interpolation, as the value returned from the model is not
a single point prediction, but a posterior probability distribution
which describes the uncertainty of the prediction, along with the
``best estimate'' prediction as the mean of $p(\vec{y}^* | \vec{x}^*, \mathcal{D})$.

\subsection{Choice of covariance function}
\label{sec:covariancefunction}

A covariance function can be designed for any given \GP{} by considering both the
hyperparameters and functional form of the covariance function.
A much fuller discussion of these considerations is given in~\cite{2016PhRvD..93f4001M}, however a summary is made here due to the importance of these considerations in the remainder of this work.

A covariance function must be positive definite, that is, it returns a value which is non-negative for any element in its domain.
Practically, when working with data, this means that the covariance function will map any pair of points in the set of data to a non-negative real number.
We can additionally require a covariance function to be stationary, in which case it is a function of $x_1 - x_2$, and so invariant to translations in the input space. Further, if it is a function of $|x_1-x_2|$ only is is isotropic, and invariant to rigid motions within the input space~\cite{Rasmussen:2005:GPM:1162254}.

A straight-forward function of $x_1 - x_2$ is a distance function of the form
\begin{equation}
  \label{eq:distancefunctioneuclid}
  d^2(x_1, x_2) = \sum_{a,b} (x_1 - x_2)^a (x_1 - x_2)^b.
\end{equation}

Such a distance function is stationary, and a covariance function using this distance metric will then be a stationary \GP{}.

The functional form of the covariance function is important in defining the prior belief about the form of the function which generated the training data.
A common choice of covariance function is the exponential squared covariance function
\cite{Rasmussen:2005:GPM:1162254},
\begin{equation}
  \label{eq:squaredexponentialkernel}
  k_{\SE}(d; \lambda) = \exp\left( - \frac{d^2}{2\lambda^2} \right).
\end{equation}
For $\lambda$ the length-scale of the kernel, which can be tuned as a hyperparameter. A larger value of this parameter will describe longer scale variations within the data.

The functional form of the squared exponential covariance function implies that the generating function was infinitely differentiable, however, generalisations of this covariance function allow the differentiability to be altered through the addition of a further hyperparameter, allowing the smoothness of the generating function to be learned during the training of the GP.

An example of such a covariance function is the general Mat\'{e}rn covariance function, which has the form
\begin{equation}
  \label{eq:matern}
  C_\nu(d; \rho, \nu) = \frac{2^{1-\nu}}{\Gamma(\nu)}\Bigg(\sqrt{2\nu}\frac{d}{\rho}\Bigg)^\nu K_\nu\left(\sqrt{2\nu}\frac{d}{\rho}\right),
\end{equation}
for $\Gamma$ the gamma function, $K_{\nu}$ the modified Bessel function of the second kind, and $\rho$ and $\nu$ are hyperparameters.
A \ac{GP} which uses this covariance function will be $(\nu-1)$-times differentiable~\cite{Rasmussen:2005:GPM:1162254}.

Uncertainty in the training data used to train the \GP{} can be accounted for by modifying the covariance matrix appropriately, with $K^{+}$ of equation~\ref{eq:blockK-plus-mat} becoming
\begin{equation}
  \label{eq:blockK-plus-mat-noise}
  K^+ =
  \begin{bmatrix}
    K_{x,x} +  \sigma_i^2 I & K_{x,x^*} \\ K_{x^*,x} & K_{x^*, x^*}
  \end{bmatrix},
\end{equation}
for $I$ the identity matrix, and $\sigma_i$ the standard deviation of the $i$-th datum.

The predictive distribution then becomes

\begin{equation}
    \label{eq:predictive-gp-noise}
\begin{aligned}
    p(\vec{y}^* | \vec{x}^*, \mathcal{D}) = \mathcal{N}(\vec{y}^* |& K_{x^*,x} (K_{x,x}+\sigma^2_iI)^{-1} \vec{y}, \nonumber\\ & K_{x^*, x^*} - K_{x^*,x}(K_{x,x}+I\sigma_i)^{-1} K_{x,x^*}).
  \end{aligned}
\end{equation}

The inclusion of a small noise term\update{, by setting $\sigma$ to a small value, such as $10^{-6}$,}  is often advantageous for improving the numerical stability of the inversion of the covariance matrix (Tikhonov regularisation), which can otherwise become nearly-singular as the total amount of training data increases.

More complex covariance models can be obtained by combining simpler covariance functions through addition or multiplication.
This allows the modelling of effects within the training data which occur at different scale lengths, or with different properties.
For example, if the training data is produced by a process with a long-term variation, but within that long-term variation there are a number of short-term variations, we might model this as a combination of two covariance functions, specifically the sum of two exponential squared covariance functions.
“Similarly, it is possible to define a \GP{} that uses different kernels in different dimensions of the parameter space, allowing the scale length of each dimension to be chosen individually; for this purpose we might use a kernel that is a product of different kernels for each dimension.
In the case of a diagonal metric this happens automatically when using the squared-exponential covariance function, and covariance functions with similar form, since they determine the scale of each dimension independently.

\subsection{Training the Surrogate}
\label{sec:training}

Then, in order to produce a good fit to the data, and to accurately estimate the uncertainty of the prediction from the regression model we performed Bayesian model selection to determine the optimal value of the covariance function's hyperparameters.
In order to initialise this process we made a rough guess of appropriate values for the hyperparameters; we do this by calculating the average distance between points along each axis in the
data space, and using this as our initial estimate \update{for the hyperparameter values}.
Starting from these initial values we optimised the log-likelihood of the model by varying the hyperparameter values to determine a maximum \textit{a posteriori} log-likelihood.

\update{In order to cope with the large number of training points, and to increase the speed of the training process we used the ADAM~\cite{2014arXiv1412.6980K} optimisation algorithm to stochastically optimise the log-likelihood using mini-batches of $100$ training points.}

While this method of determining, and fixing, the hyperparameters of the \GP{} is computationally convenient, other methods are also possible, including marginalising over the hyperparameters.
However, our method has the advantage that it is not necessary to evaluate the \ac{GPR} model for all of the hyperparameter samples, and can therefore be evaluated more rapidly.

\section{The \HERON{} model}
\label{sec:heron}

Using a \ac{GPR} model, named \HERON{}, trained on \ac{NR} data from the Georgia Tech \ac{BBH} waveform catalogue.
\HERON{} was designed as a surrogate model operating over the eight dimensions of the \ac{BBH} parameter space, and we present it as a proof-of-concept of a \ac{GPR}-based surrogate for this larger parameter space.
The model is constructed using a squared-exponential covariance function.
We will demonstrate that this model is capable of producing waveforms for spinning and precessing \ac{BBH} systems.

\subsection{Training Data}
\label{sec:trainingdata}

We constructed our training data for the \HERON{} model from the strain values of the \numbertrainingwaveforms{} waveforms in the Georgia Tech Catalogue~\cite{2016CQGra..33t4001J}.
These data were acquired in the LIGO Numerical relativity \texttt{hdf5} format~\cite{2017arXiv170301076S}, and the \texttt{pycbc} package~\cite{2016CQGra..33u5004U,alex_nitz_2017_883086,2014PhRvD..90h2004D} was used to produce the $(2,2)$-mode of these waveforms.

\begin{figure*}
  \centering
  \includegraphics[width=\textwidth]{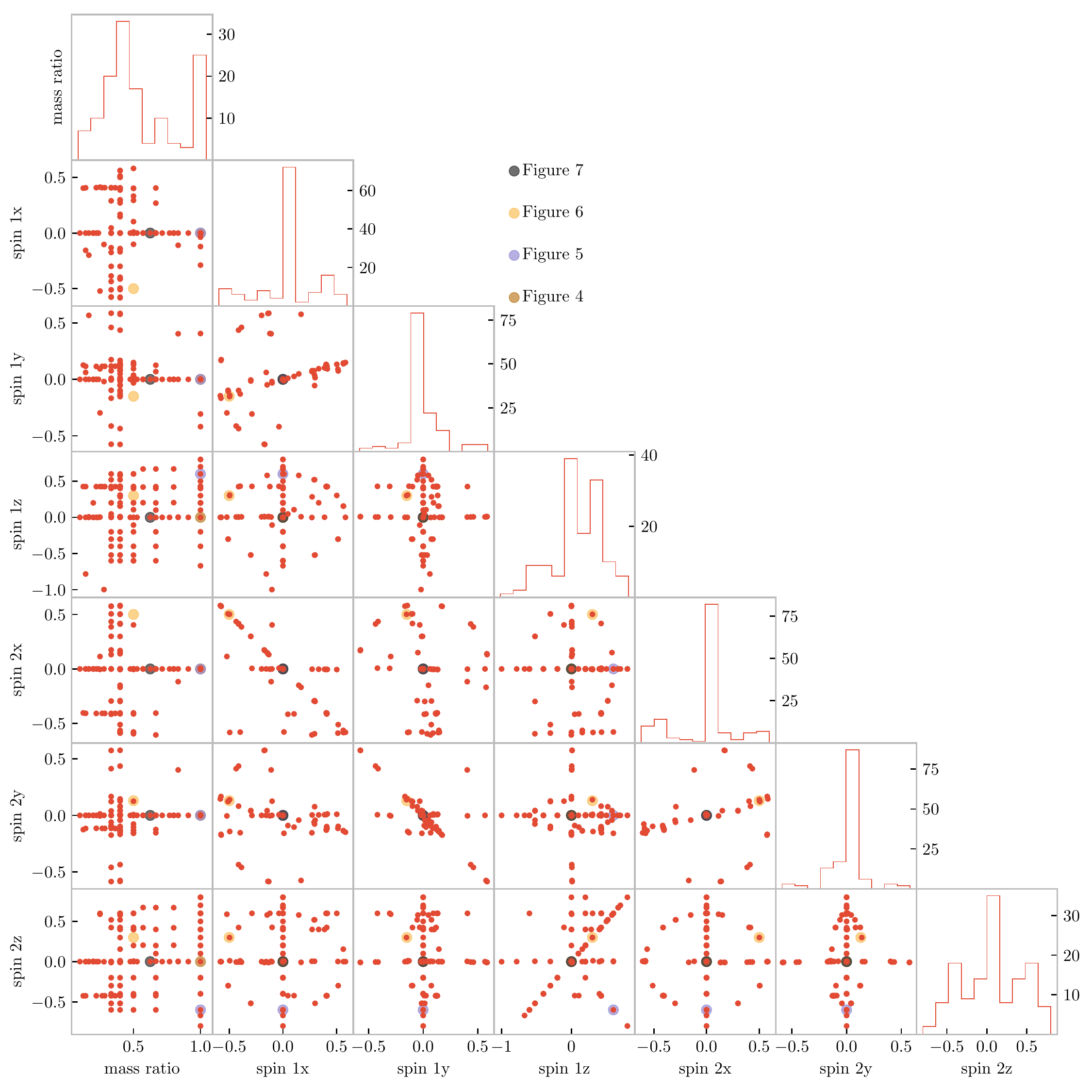}
  \caption{A pair-plot of the parameter space sampling in the Georgia Tech catalogue.
    The subplots on the diagonals show the histograms of the distribution of waveforms (as red points) generated with respect to each individual parameter.
    Three additional points are displayed on the plot corresponding to the waveform samples shown in the later figures of this paper.
  }
  \label{fig:pairplotparams}
\end{figure*}

Each waveform is parameterised by seven quantities (the mass ratio and the spin vectors of each component black hole) in a vector we denote $\vec{x_i}$.
Each strain value, $h_i$, within the waveform is further parameterised by a time relative to the maximum strain value in the waveform, and thus each training point is parameterised by an 8-dimensional parameter vector, which we denote $\vec{x_i}'$.
This provides us with a training set which has 8 input dimensions, and a single output dimension, with the form
\[ \set{D} = \left\{ (\vec{x'_i},  h_i) | i = 1, 2, \dots, N \right\} \]
for $N$ the total number of strain samples used from all of the training waveforms.
The distribution of training waveforms throughout the parameter space is shown in figure~\ref{fig:pairplotparams}.

\subsection{Computational Complexity}
\label{sec:complexity}

A major drawback of GPR is the need to invert the covariance matrix in order to produce predictions.
Matrix inversion is a computationally intensive task which scales in memory with $N^2$, for $N$ training points, and with $N^3$ in time.
The standard approach to GPR described in equation~\ref{eq:predictive-gp} thus rapidly becomes impractical, requiring large quantities of memory for even moderately sized training sets. 
In order to overcome these scaling problems, approximate \acp{GP} simplify the inversion of the covariance matrix by making simplifying assumptions about its form.
One example is the use of the approximate \ac{HODLR}~\cite{hodlr} inversion method, 
which allows inversion to be carried out in $\mathcal{O}(N \log^2 N)$ operations.
This approach is possible because kernels such as the exponential squared kernel produce covariance matrices which can be arranged to form  \ac{HODLR} matrices.
The off-diagonal blocks are then factorised using partial-pivoted LU decomposition, and the on-diagonal blocks are factorised using a more accurate algorithm, such as Cholesky decomposition.
The block inverses are then recombined to provide the (approximate) overall matrix inverse.

\update{This surrogate model makes use of $N = 4,740$ training points, stored as 64-bit floating points, and requiring approximately $370$ kilobytes to store in memory.
  This leads to the need to invert a covariance matrix which requires around $134$ gigabytes of memory.
  To overcome this} we employed the HODLR method for calculating the matrix inverse, using the implementation in the \texttt{George}~\cite{hodlr} Python package.

\update{The use of an approximate method to produce the \ac{GP} posterior will introduce additional uncertainties. While tests conducted in~\cite{hodlr} indicate that this additional uncertainty is likely to be small, we make use of in-sample testing (see section \ref{sec:tests:insample}) to assess the impact of using this method on the model's ability to replicate its training data.}

\section{Verification of the \ac{GPR} model}
\label{sec:verification}

The sparsity of training data poses a considerable challenge to the testing and verification of a model such as the \HERON{} model; conventional approaches to testing such a model involve setting aside a fraction of the training data to compare to the model output when evaluated at the parameter space location of each test datum. 

The quantity of numerical relativity waveforms available at present in the Georgia Tech catalogue makes this approach difficult, as some regions of the parameter space are very sparsely sampled, and omitting a training waveform in this location may significantly complicate the process of training the model.
To overcome this we have carried out three separate categories of test on the \HERON{} model.

\begin{description}
\item[In-sample tests] where the entire catalogue of available training waveforms are used to condition the \ac{GPR} used by the model. Waveforms are then produced from the model at the parameter locations which correspond to each of the training waveforms, and the match between the \HERON{} waveform and the \ac{GPR} waveform is calculated.

\item[Out-of-sample tests] where a single waveform from the catalogue is omitted from the set of training waveforms used to condition the \ac{GP}. A \ac{GPR} model is conditioned on a reduced catalogue for each waveform, the model is retrained to find the optimal hyperparameters given the reduced dataset, and the waveform is produced from the reduced \HERON{} model which corresponds to the omitted \ac{NR} waveform. The match is then computed between these two waveforms.

\item[Tests against phenomenological models] where the match is computed between waveforms produced by \HERON{} and by other waveform models, such as \texttt{SEOBNRv3} and \texttt{IMRPhenomPv2}.
\end{description}

Each approach to testing has different advantages and disadvantages, and test for different aspects of the model's performance.

For each of the tests we compare the output of the \HERON{} model with another waveform by calculating the mismatch between the two waveforms.
This is defined as 
\begin{equation}
  \label{eq:mismatch}
  \mathcal{M}(h_{\text{model}}, h_{\text{ana}}) = 1 - \max_{t_0, \phi_0} \frac{ \langle h_{\text{model}}, h_{\text{ana}} \rangle}
  {\sqrt{ \langle h_{\text{model}}, h_{\text{model}} \rangle \langle h_{\text{ana}}, h_{\text{ana}} \rangle}}.
\end{equation}
where $h_{\text{model}}$ and $h_{\text{ana}}$ are respectively the timeseries predicted by the \ac{GPR} model and the analytical phenomenological approximant, $t_0$ and $\phi_0$ are the merger time and merger phase, and $\langle \cdot, \cdot \rangle$ is the noise-weighted inner product between two waveforms, defined as
\begin{equation}
  \label{eq:noiseweightedinner}
  \langle a, b \rangle = \Re \int_{- \infty}^{\infty} \frac{ \tilde{a}^*(f) \tilde{b}(f) }{ S_n (f) } \dd{f} 
\end{equation}
for $\tilde{a}$ and $\tilde{b}$ respectively the Fourier transforms of the timeseries $a$ and $b$, $S_n$ the amplitude spectral density of the noise, and $f$ the frequency.

For all of the tests presented in this paper we assume that the noise is flat across frequencies, that is \update{$S_n (f) = 1 \forall f$}.

\subsection{In-sample tests}
\label{sec:tests:insample}

The simplest set of tests which we perform on the \HERON{} model are \emph{in-sample} tests, which effectively test the model's ability to reproduce its own training data. 
For the \HERON{} model this involved computing the mean waveform from the \ac{GP} corresponding to each waveform which was used in the training set.
The mismatch was then calculated between each mean waveform and the corresponding \ac{NR} training waveform using the expression for waveform mismatch, $\mathcal{M}$, given in equation \ref{eq:mismatch}.

\begin{figure*}
\includegraphics{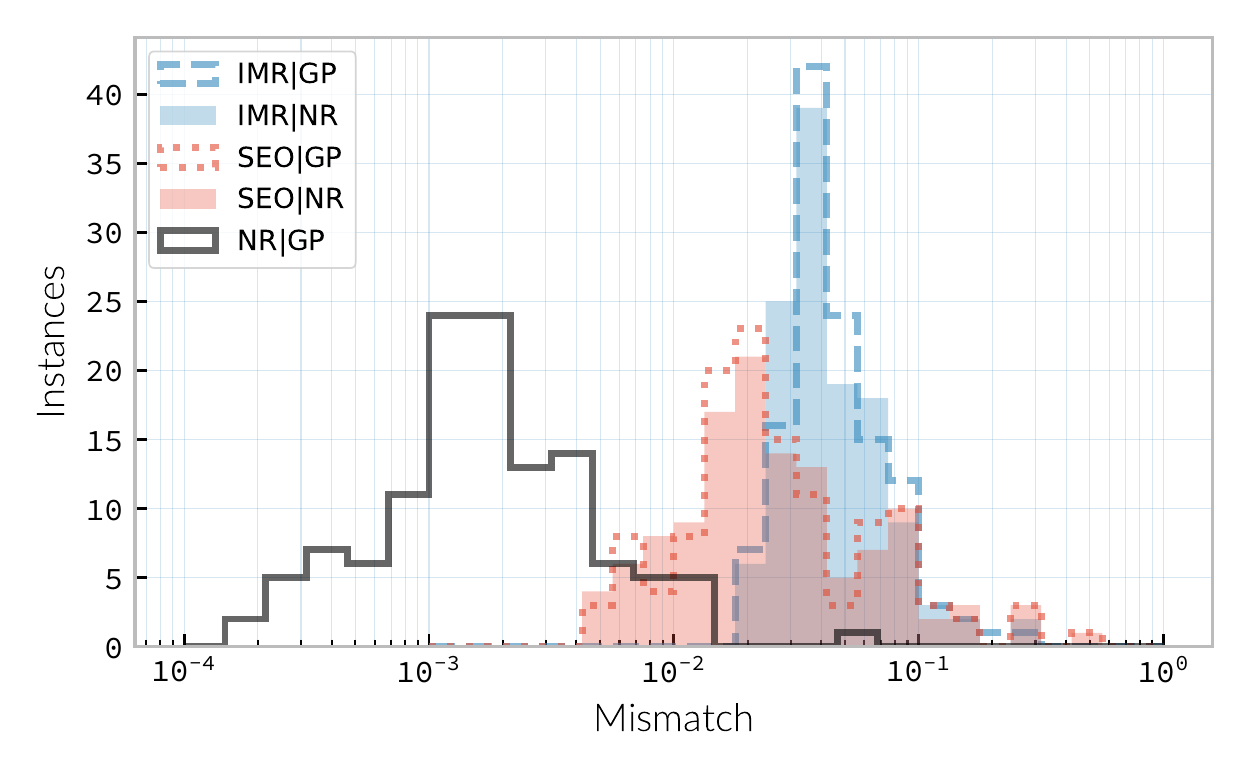}
\caption[Mismatches between \HERON{}, Georgia Tech waveforms, and phenomenological approximants from in-sample testing]{The distributions of mismatches between waveforms from the \HERON{} model and each of the \ac{NR} waveforms from the Georgia Tech waveform catalogue (black outline histogram) used in the training set using the procedure described in section \ref{sec:tests:insample}.
Additionally, the mismatch distributions between waveforms produced at the same parameters as the \ac{NR} waveforms by the \texttt{SEOBNRv3} (red outline histogram), and the \texttt{IMRPhenomPv2} (blue outline histogram) phenomenological waveform models are plotted. 
For comparison the distributions of mismatch between the same Georgia Tech waveforms and the corresponding waveforms from the \texttt{SEOBNRv3} and \texttt{IMRPhenomPv2} models are plotted as filled red and blue histograms respectively.
 \label{fig:tests:in:hist}}
\end{figure*}

In-sample testing ought to reveal problems with the choice of hyperparameters in the model, inconsistencies in the training data itself\update{, and error introduced into the model through the use of an approximate method for the inversion of the covariance matrix}.
Figure  \ref{fig:tests:in:hist} plots the histogram of the mismatch (equal to $1-\mathcal{M}$) values which resulted from these tests against the Georgia Tech waveforms used as the training data (plotted as the black-outlined histogram).
Reassuringly the mismatch between the vast majority of the model outputs and the training data are small.
\update{The mean mismatch from these in-sample tests is $0.003$, with $95\%$ of the mismatches falling between $0.000245$ and $0.0124$.
This implies that the additional error introduced into the waveform using the approximate matrix inversion technique is responsible for only a small mismatch when compared to the \ac{NR} waveform.  
}

\subsection{Out-of-sample tests}
\label{sec:tests:outsample}

A more rigorous test of a predictive model involves comparing the model's output in a region of the parameter space which does not contain a training datum. This process, known as out-of-sample testing, is difficult for the \HERON{} model, thanks to the large (seven dimensional) parameter space, and the small number of available training waveforms.
As a result, removing a substantial fraction of the waveforms in order to produce a set of test data would be likely to substantially affect the predictive power of the model.

\begin{figure*}
\includegraphics{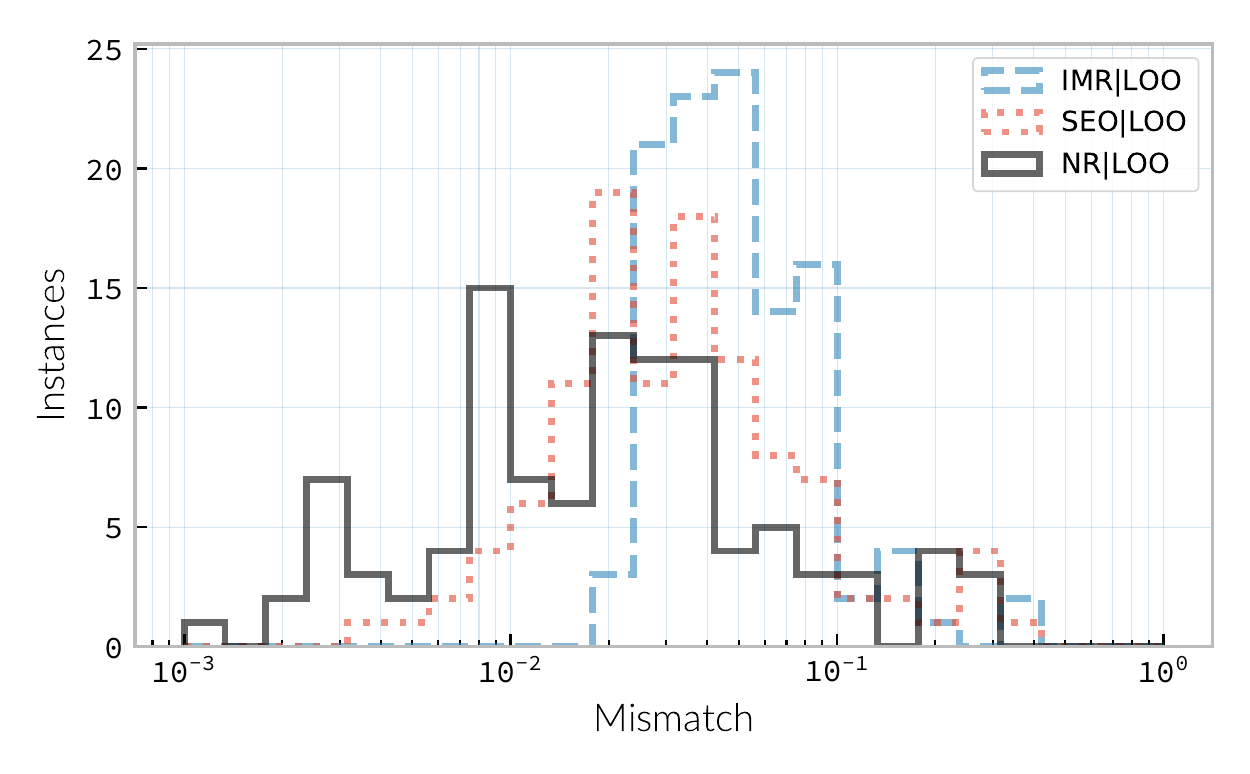}
\caption[Mismatches between \HERON{}, Georgia Tech waveforms, and phenomenological approximants from leave-one-out sampling]{The distributions of mismatches between waveforms from the \HERON{} model and each of the \ac{NR} waveforms from the Georgia Tech waveform catalogue (black outline histogram) used in the training set using the \ac{LOO} testing procedure detailed in section \ref{sec:tests:outsample}.
Additionally, the mismatch distributions between waveforms from the \HERON{} model and  waveforms produced at the same parameters as the \ac{NR} waveforms by the \texttt{SEOBNRv3} (red outline histogram), and the \texttt{IMRPhenomPv2} (blue outline histogram) phenomenological waveform models are plotted. 
 \label{fig:tests:out:hist}}
\end{figure*}

To overcome this we performed a \ac{LOO} testing procedure.
In order to do this multiple training datasets are produced; from each a single waveform is omitted.
This reduced dataset is then substituted for the data on which the full \HERON{} model's \ac{GP} is conditioned, \update{and the model is retrained using the reduced training set, in order to find the hyperparameter values which maximise the model's log-likelihood}.
The reduced \HERON{} model is then evaluated at the parameter location corresponding to the omitted waveform, in order to compute a predicted mean waveform. 
The mismatch between the predicted waveform and the omitted \ac{NR} waveform was then computed, and the distribution of these mismatches is plotted in figure \ref{fig:tests:out:hist} as a black-outlined histogram.

\update{The mean mismatch across all of the tests was $0.0369$, with $95\%$ of the mismatches between $0.000922$ and $0.226$.
  A total of 8 tests produce a mismatch greater than $0.1$, and in each case the variance of the returned waveform is very large, indicating that the model is able to express its lack of knowledge about these regions of the parameter space effectively.
  While this uncertainty could be directly incorporated into some applications of the model it could also be used to automatically flag draws from the model which are of low confidence, and which should not be relied upon in an analysis.
  }

\subsection{Tests against phenomenological models}
\label{sec:tests:phenom}

It may also be helpful to understand how the outputs of the \HERON{} model compare to conventional phenomenological approximants which are in widespread use.
To do this we calculated the mismatch between the output of the \HERON{} model at the same parameter locations as the in-sample and leave-one-out tests.

In the left panel of figures \ref{fig:nonspin-equalmass} and \ref{fig:spinanti-equalmass}, we compare the waveform computed for different random samples drawn from the \ac{GPR} model, the mean of the \ac{GPR} model and the \texttt{IMRPhenomPv2} and \texttt{SEOBNRv3} waveforms for a non-spinning configuration (figure \ref{fig:nonspin-equalmass}), an equal-mass configuration with anti-aligned spins (figure \ref{fig:spinanti-equalmass}), and a precessing configuration (figure \ref{fig:prec-equalmass}).
The distribution of mismatches between the \ac{GPR} model predictions and the two phenomenological approximants are shown in the right panel of each figure, with matches calculated between the approximant waveforms and one-hundred sample waveforms drawn from the \ac{GPR} model.
In addition, the mismatch between the mean waveform produced by the \ac{GPR} model and each phenomenological model is indicated by a solid line; it is noteworthy that this mismatch is always smaller than the mean of the mismatches between the sample draws and the phenomenological models.
This is a result of the mismatch being a somewhat asymmetric indicator: the mismatch will always be higher for a waveform which over-estimates or under-estimates some feature of the waveform, where the over- and under-estimates will be averaged through the use of the mean waveform, producing a lower mismatch.

In the in-sample case the \HERON{} model reproduces the \ac{NR} waveforms with substantially lower mismatch than either phenomenological model. 
This behaviour is to be expected, since the \HERON{} model has direct access to the \ac{NR} data, where the phenomenological models do not. 
It is worth noting that the mismatch for \texttt{SEOBNRv3} is consistently smaller than that of \texttt{IMRPhenomPv2} against both \ac{NR} and the \HERON{} model. 
\texttt{IMRPhenomPv2} is known to be accurate over a smaller range of black hole spins than the \texttt{SEOBNRv3} model.

We also compare the behaviour of the \ac{LOO} models described in section \ref{sec:tests:outsample} with the two phenomenological models.  The distributions of mismatches from comparison between waveforms from the \ac{LOO} models and waveforms produced by each approximant at the same parameter location as the \ac{NR} waveform which was omitted from the \ac{LOO} model, are plotted in figure \ref{fig:tests:out:hist} as blue and red-outline histograms for the \texttt{IMRPhenomPv2} and \texttt{SEOBNRv3} waveforms respectively.  Here we see that the \ac{LOO} models are generally in good agreement with the two approximants, with the mismatches slightly larger between the \ac{LOO} models and the approximants than between the \ac{LOO} models and the \ac{NR} waveforms, which is also seen in the in-sample testing.

\section{Example Waveforms}
\label{sec:examples}

While we have discussed at length the various tests which we carried out on the \HERON{} model, it is valuable to be able to visually compare the output of this model with the phenomenological models used in testing.

\begin{figure*}

  \includegraphics[width=\textwidth]{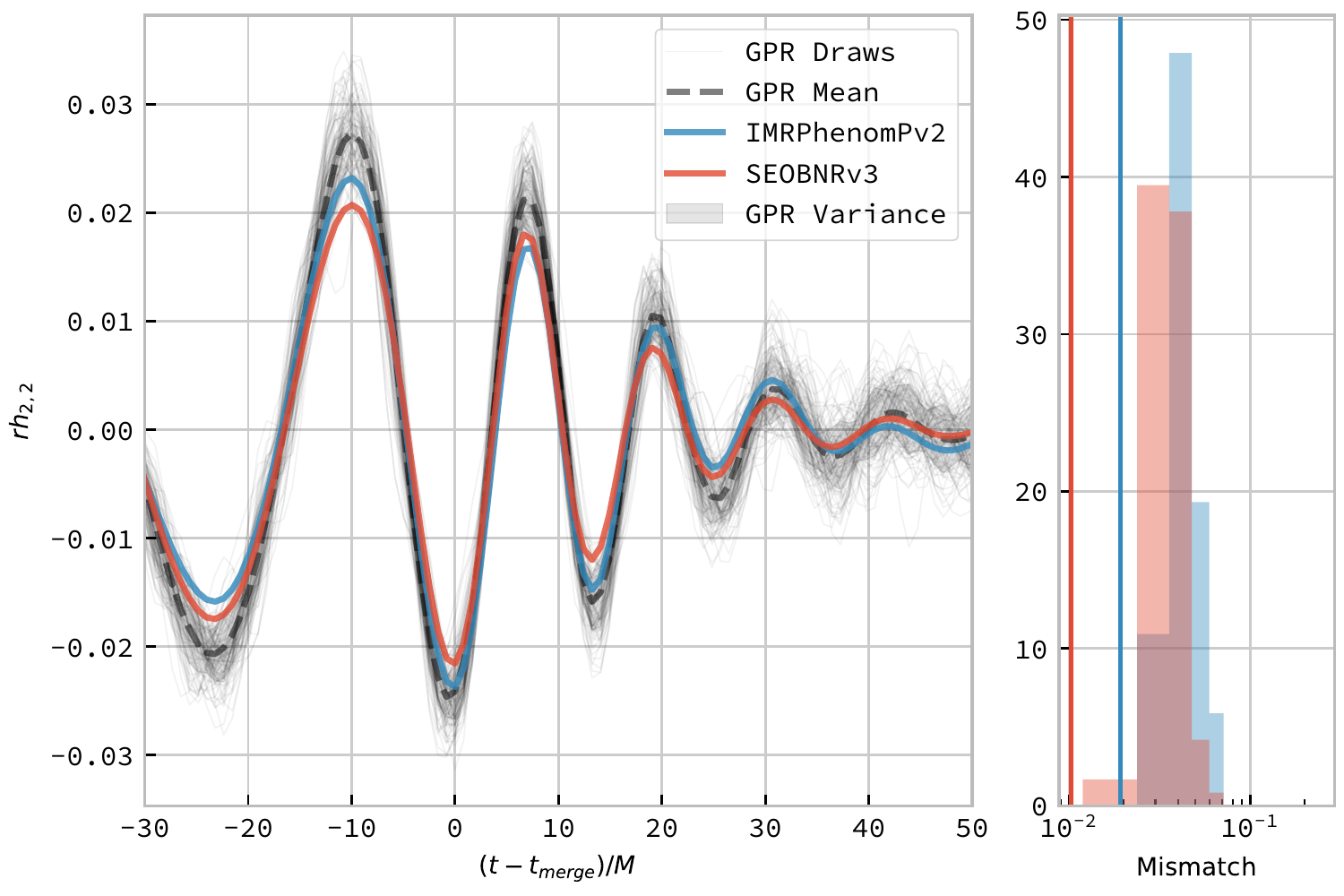}
  \caption{\textbf{Non-spinning waveform}. One hundred draws from the Gaussian process (left panel) for a non-spinning, equal-mass configuration ($\vec{s_1} = (0,0,0)$, $\vec{s_2} = (0,0,0)$, $\vec{q} = 1.0$), shown as light grey lines compared to two analytical phenomenological approximant models, \texttt{SEOBNRv3} and \texttt{IMRPhenomPv2} in red and blue respectively. The mean draw from the Gaussian process is shown as a grey dashed line, while the associated variance is plotted as a grey-filled region surrounding the mean. In the right panel the distribution of mismatches between the samples and both phenomenological waveforms are shown, with the vertical lines representing the mismatch between the mean waveform from the \ac{GPR} and the phenomenological waveform.
    \label{fig:nonspin-equalmass}
  }
\end{figure*}

\begin{figure*}
  \includegraphics[width=\textwidth]{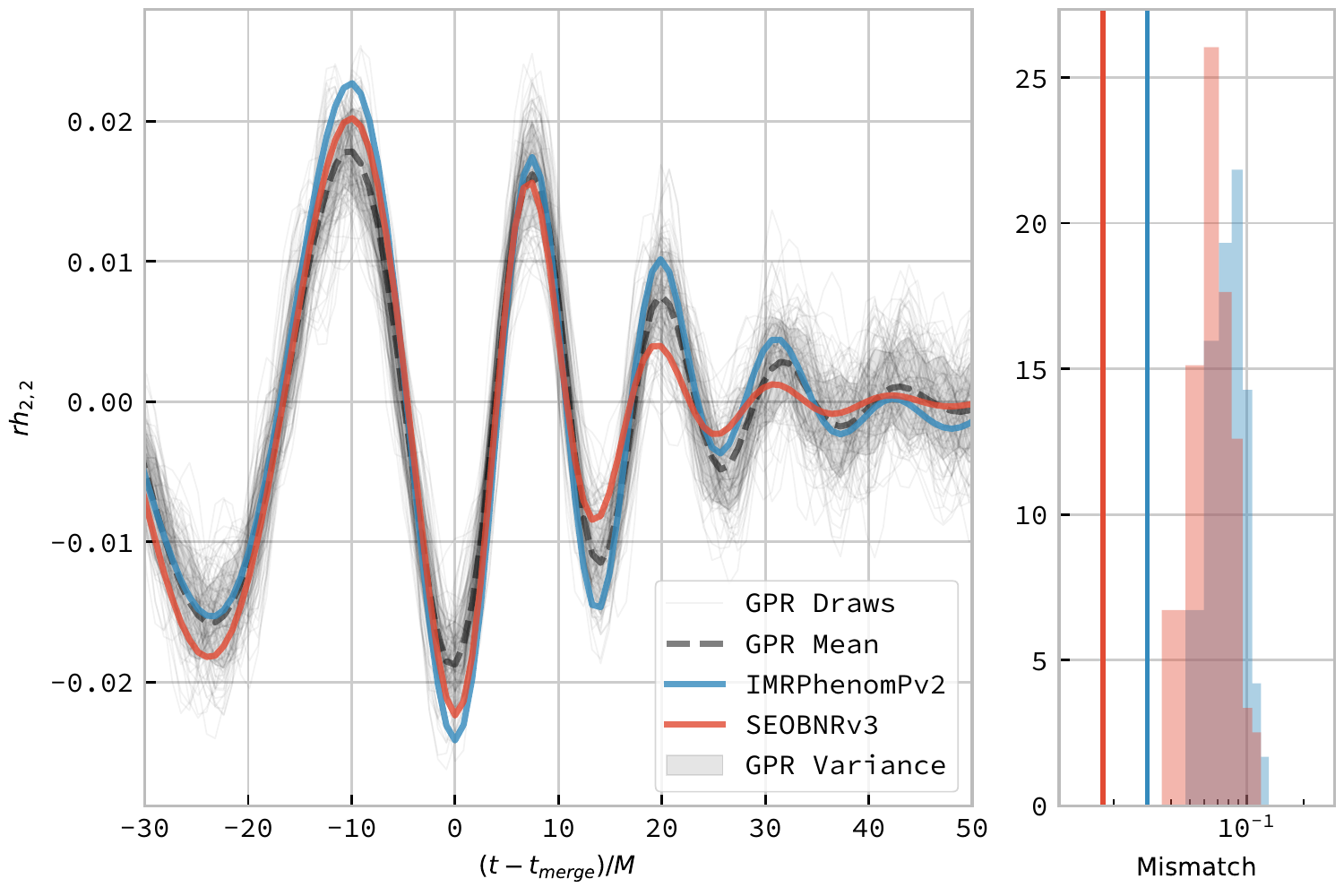}
  \caption{\textbf{Anti-aligned spin waveform}. One hundred draws from the Gaussian process (left panel) for a non-spinning, equal-mass configuration ($\vec{s_1} = (0,0,0.6)$, $\vec{s_2} = (0,0,-0.6)$, $\vec{q} = 1.0$), shown as light grey lines compared to two phenomenological approximant models, \texttt{SEOBNRv3} and \texttt{IMRPhenomPv2} in red and blue respectively. The mean draw from the Gaussian process is shown as a grey dashed line, while the associated variance is plotted as a grey-filled region surrounding the mean. In the right panel the distribution of mismatches between the samples and both phenomenological waveforms are shown, with the vertical lines representing the mismatch between the mean waveform from the \ac{GPR} and the phenomenological waveform.
    \label{fig:spinanti-equalmass}
  }
\end{figure*}

In the left panel of figures~\ref{fig:nonspin-equalmass} and~\ref{fig:spinanti-equalmass}, we compare the waveform computed for different random samples from the \ac{GPR} model, the mean of the \ac{GPR} model and the \texttt{IMRPhenomPv2} and~\texttt{SEOBNRv3} waveforms for a non-spinning configuration (figure~\ref{fig:nonspin-equalmass}) and an equal-mass configuration with anti-aligned spins (figure~\ref{fig:spinanti-equalmass}).
The distribution of mismatches between the \ac{GPR} model predictions and the two phenomenological approximants are shown in the right panel of each figure, with matches calculated between the approximant waveforms and one-hundred sample waveforms drawn from the \ac{GPR} model.

\begin{figure*}
  \includegraphics[width=\textwidth]{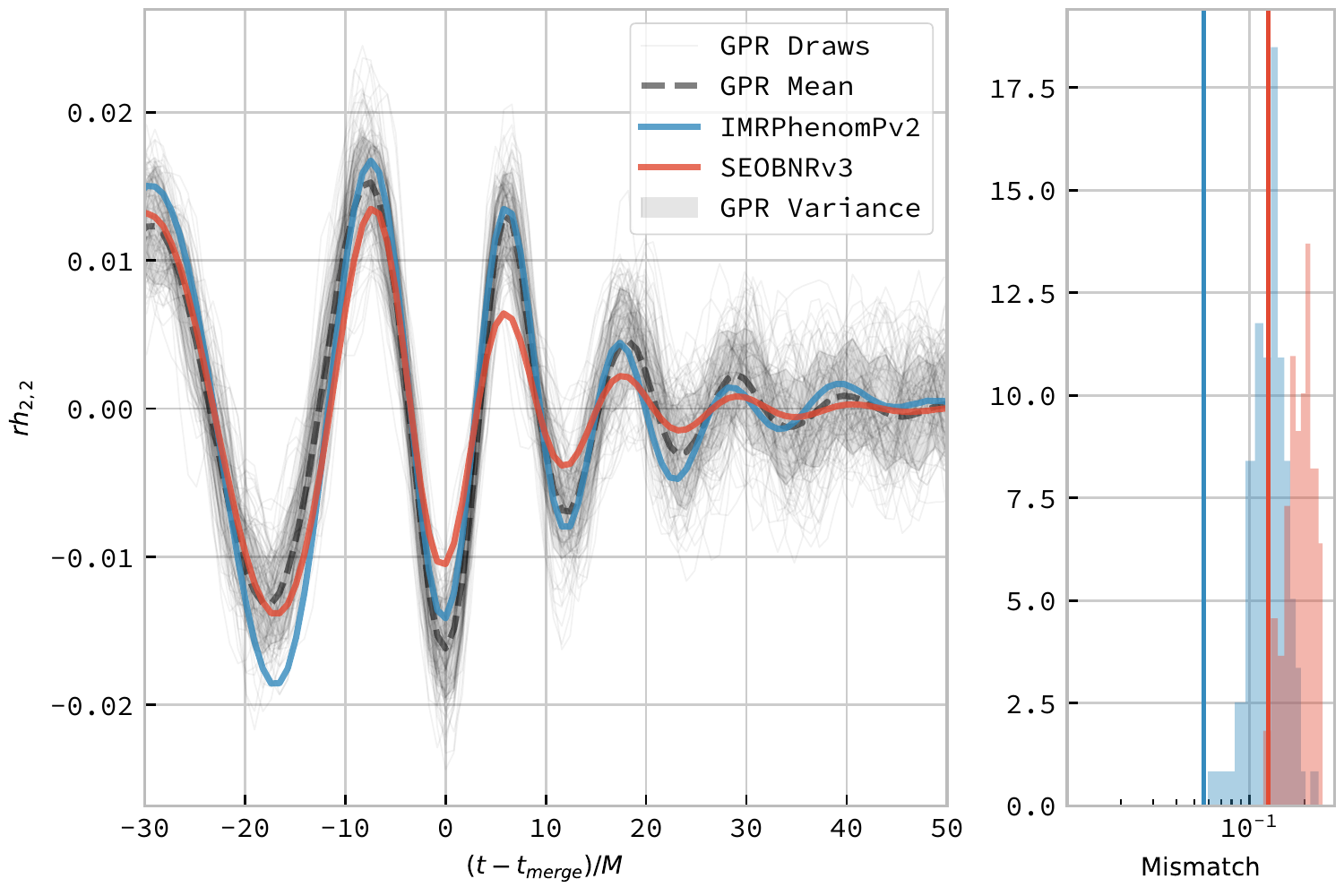}
  \caption{\textbf{Precessing waveform}. One hundred draws from the Gaussian process (left panel) for a precessing system, with a mass ratio $q=0.4$, and a spin configuration ($\vec{s_1} = (-0.5,-0.15,0.3)$, $\vec{s_2} = (0.5, 0.13, 0.3)$), shown as light grey lines compared to two phenomenological approximant models, \texttt{SEOBNRv3} and \texttt{IMRPhenomPv2} in red and blue respectively. The mean draw from the Gaussian process is shown as a grey dashed line, while the associated variance is plotted as a grey-filled region surrounding the mean. In the right panel the distribution of mismatches between the samples and both phenomenological waveforms are shown, with the vertical line representing the mismatch between the mean waveform from the \ac{GPR} and the phenomenological waveform.
    \label{fig:prec-equalmass}
  }
\end{figure*}

An example of a precessing waveform generated by the \ac{GPR} model is also shown in figure~\ref{fig:prec-equalmass}.

\begin{figure*}
  \includegraphics[width=\textwidth]{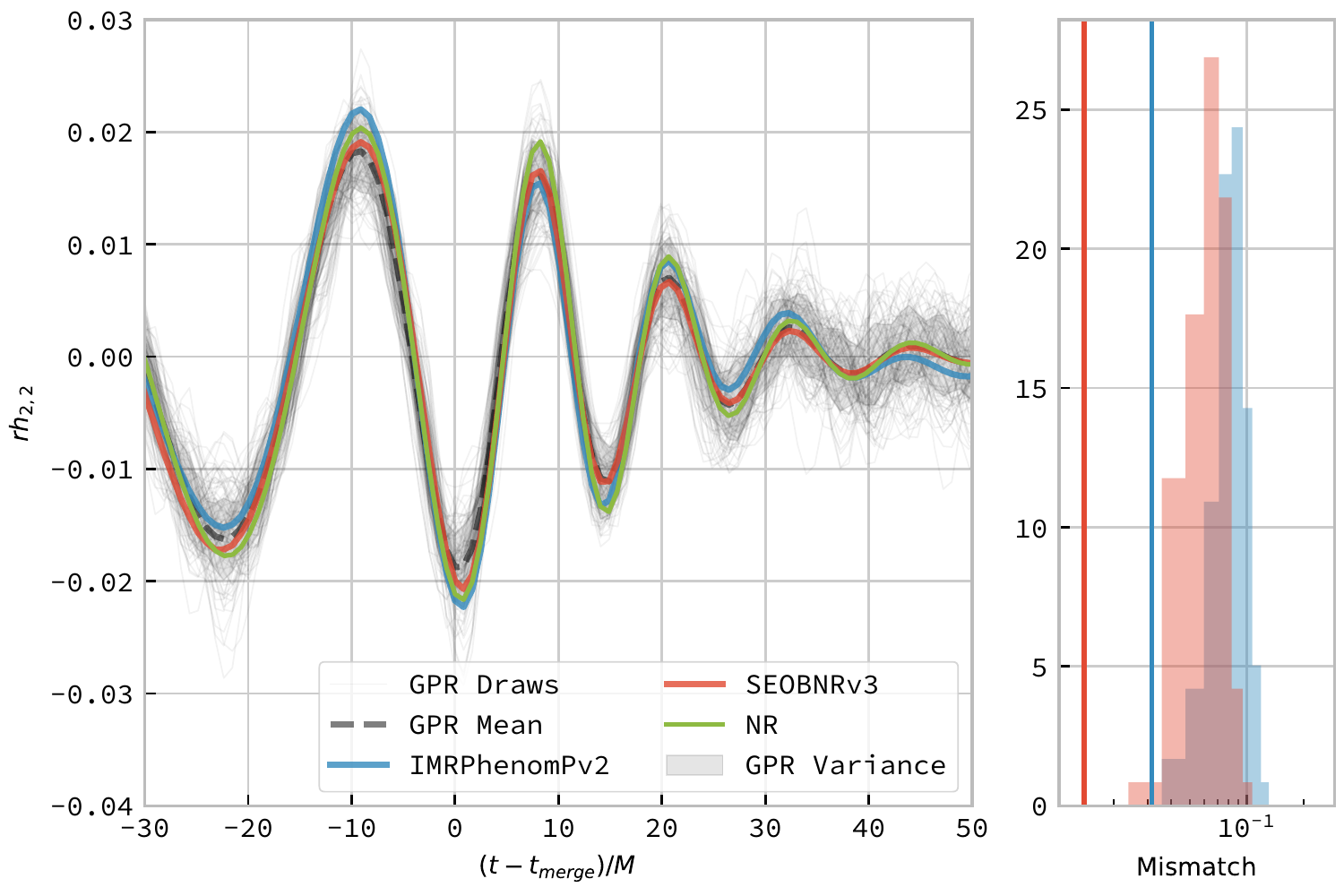}
  \caption{\textbf{\ac{GPR} predictions, compared to \ac{NR}}. One hundred draws from the Gaussian process (left panel) for a non-spinning configuration ($\vec{s_1} = (0,0,0)$, $\vec{s_2} = (0,0,0)$, $\vec{q} = 0.625$), \update{shown as light grey lines compared to the phenomenological approximant models, \texttt{IMRPhenomPv2} in blue; and \texttt{SEOBNRv3} in red}. The mean draw from the Gaussian process is shown as a grey dashed line, while the associated variance is plotted as a grey-filled region surrounding the mean. 
The differences between the phenomenological model and the \ac{GPR} model waveforms are seen to also exist between the phenomenological model waveforms and the \ac{NR}-derived waveform, plotted here in purple. In the right panel the distribution of mismatches between the samples and both phenomenological waveforms are shown, with the vertical lines representing the mismatch between the mean waveform from the \ac{GPR} and \update{each} phenomenological waveform.
    \label{fig:prediction-with-nr}
  }
\end{figure*}

In figure~\ref{fig:prediction-with-nr} we also show one of the training \ac{NR} waveforms plotted alongside the mean output of the \ac{GPR} model, one-hundred waveform draws from the model, and waveforms produced from both of the phenomenological models used for the comparisons in figures~\ref{fig:nonspin-equalmass} to~\ref{fig:prec-equalmass}.

\section{Summary}
\label{sec:discussion}

We have entered the era of routine \ac{GW} detection, and the ability to accurately and rapidly characterise signals from events such as \ac{BBH} coalescences will be critical to understanding the properties of these systems.
This characterisation process relies on the availability of waveform templates which are either precomputed prior to the analysis being run, or can be generated on-the-fly.
Highly accurate waveforms, generated by \ac{NR} simulations, are able, and in principal can facilitate accurate inference on detected signals.
However, the expense of producing them limits their coverage of the parameter space; as a result of this lack of coverage, and the considerable time requirements to produce new waveforms, any inference method which relied solely on \ac{NR} techniques could not hope to satisfy the requirement to rapidly characterise signals, and would not be practical in a scenario where multiple events are detected every month.
Phenomenological models, which can be evaluated rapidly, are available, which attempt to interpolate across a large volume of the parameter space, but the accuracy of the waveforms which they produce can be difficult to assess.
This leads to the possibility of introducing biases into the inferred properties of the system which generated the signal.

In this paper we have laid-out an approach to improving the accuracy of gravitational wave parameter estimation in the context of limited template availability by implementing a waveform approximant model using \ac{GPR}, providing not only a point-estimate of the waveform at any point in the \ac{BBH} parameter space, but also a distribution of plausible waveforms, allowing the uncertainty of the interpolation to be taken into account during the analysis.
In contrast to previous attempts to produce a \ac{GPR} model for \ac{GW} waveforms, such as~\cite{2017zoheyrsurrogate}, our model is trained on data from the Georgia Tech \ac{NR} waveform catalogue, described in section~\ref{sec:trainingdata}. 

We introduced \ac{GPR} in section~\ref{sec:gps} as a non-parametric regression method.
This property allows the regression model to be constructed while making minimal assumptions about the form of the waveforms, which are encoded through the form of the covariance function.
We discuss covariance functions in section \ref{sec:covariancefunction},
In order to reduce the computational burden of evaluating the model a hierarchical matrix inversion method was used (described in \cite{hodlr} and discussed in section \ref{sec:complexity}).

We present three testing strategies for our \ac{GPR} model, in addition to a number of waveforms which have been produced by it in section~\ref{sec:verification}. We present both the results of these tests, and make comparisons between the model's output and two well-established phenomenological models.
This difference also occurs between the phenomenological model and the waveform produced from \ac{NR}. A number of phenomena are likely to have contributed to this discrepancy.
One such difference in the systematic errors of the \ac{NR} simulations used to produce the training data for the \ac{GPR} model compared to those used to calibrate the phenomenological models.
Additionally, the relatively small number of waveforms used to calibrate the phenomenological models compared to the \ac{GPR} model are likely to introduce systematic errors in the waveforms produced by those models.
In order to reduce the effect of systematic errors from \ac{NR} a larger model could include waveforms from a number of different \ac{NR} waveform catalogues, however the addition of more waveforms will increase the memory requirements to both train and evaluate the model.
Our waveform model tends towards producing conservative estimates of the waveform, this is clearly visible in the variance of the precessing waveform in figure~\ref{fig:prec-equalmass}.
The use of additional waveforms is likely to improve the confidence of the model's prediction.

In order for a \ac{GPR}-based approach such as this to be practical for parameter estimation studies using data from LIGO or Virgo it would be necessary to have a means of producing waveforms which are capable of modelling a greater amount of the inspiral than our model can currently provide.
One potential approach to solving this problem is hybridising the output waveform from our \ac{GPR} model with waveforms produced from a post-Newtonian approximant, in a similar manner to that used by~\cite{2018arXiv181207865V}.
\update{This would allow us to overcome the need for much longer waveforms to be used in the training set, while still allowing the production of waveforms with lengthier inspirals than our model is currently capable of.}

\update{We note that in this work we have not attempted to benchmark this model, and compare the times required to produce sample waveforms from it compared to the analytical approximates which are currently in regular use.
  We expect to address this short-coming in future work, but acknowledge that a number of optimisations may be made to allow the model to produce results more expediently without impacting on its accuracy.}

In conclusion, we have demonstrated that \ac{GPR} is capable of being used as an interpolant for \ac{BBH} waveforms, trained directly off data from \ac{NR} simulations. While this method cannot hope to produce waveforms with the same precision as \ac{NR} itself, it does account for the uncertainty introduced through interpolation, a feature which is valuable for preventing the introduction of bias in a \ac{PE} analysis.

\section{Acknowledgements}
\label{sec:ack}

The authors wish to thank Christopher Moore, Sebastian Khan, and Vijay Varma for their insightful comments and suggestions on an earlier draft of the manuscript, and also our \update{two} anonymous reviewer\update{s} for their suggestions.

DW received support from Science and Technology Facilities Council (STFC) grants ST/N504075/1 and ST/L000946/1.
ISH is supported by STFC grant ST/L000946/1.
BK acknowledges support from NSF awards PHY-1806580, PHY-1809572, and 1333360.
This document has been assigned LIGO document reference LIGO-P1800128.

\update{The software described in this paper is available on \texttt{GitHub} (\url{https://www.github.com/transientlunatic/heron}) and on Zenodo \cite{zenodo:heron}.}
\update{The analysis and software development described in this paper was made possible thanks to a number of open source \texttt{Python} libraries, including \texttt{numpy}~\cite{software:numpy}, \texttt{matplotlib}~\cite{software:matplotlib}, and \texttt{george}~\cite{hodlr}.}

\bibliography{bibliography,nr}

\end{document}